\documentclass[aps,pre,twocolumn,showpacs,superscriptaddress,groupedaddress]{revtex4-1}

\usepackage{graphicx}
\usepackage{dcolumn}
\usepackage{bm}

\usepackage{amssymb}
\usepackage{epsfig}
\usepackage{amsmath}
\usepackage{subfigure}
\usepackage{textcomp}
\usepackage{url}
\usepackage{float}
\usepackage{color}
\usepackage{cases}
\usepackage[normalem]{ulem}

\usepackage[colorlinks=true, urlcolor=blue, anchorcolor=blue, citecolor=blue,filecolor=blue,linkcolor=blue,menucolor=blue
]{hyperref}

\newcommand{\pp}{\mathcal{\partial}}

\newcommand{\average}[1]{\left< #1 \right>}

\begin{document}

\preprint{APS/123-QED}

\title{Fluctuations and first-passage properties of
systems of Brownian particles with reset}

\author{Ohad Vilk $^{a,b,c}$}
\email{ohad.vilk@mail.huji.ac.il}
\author{Michael Assaf $^{a}$}
\email{michael.assaf@mail.huji.ac.il}
\author{Baruch Meerson $^{a}$}
\email{meerson@mail.huji.ac.il}
 \affiliation{$^a$ Racah Institute of Physics, The Hebrew University of Jerusalem, Jerusalem 91904, Israel,}
 \affiliation{$^b$ Movement Ecology Lab, Department of Ecology, Evolution and Behavior, Alexander Silberman Institute of Life Sciences, Faculty of Science, The Hebrew University of Jerusalem, Jerusalem 91904, Israel,}
 \affiliation{$^c$ Minerva Center for Movement Ecology, The Hebrew University of Jerusalem, Jerusalem 91904, Israel}


\begin{abstract}
We study, analytically and numerically, stationary fluctuations in two models involving  $N$ Brownian particles undergoing stochastic resetting in one dimension. We start with the well-known reset model where the particles reset to the origin independently (model A).
Then we introduce nonlocal interparticle correlations by postulating that only the particle farthest from the origin
can be reset to the origin (model B).   At long times models A and B approach nonequilibrium steady states. In the limit of $N\to \infty$,
the steady-state particle density in model A has an infinite
support, whereas in model B it has a compact support, similarly to the recently studied ``Brownian bees" model.  A finite system radius, which scales at large $N$ as
$\ln N$, appears  in model A when $N$ is finite.
In both models we study stationary fluctuations of the center of mass of the system and of the system's radius  due to the random character of the Brownian motion and of the resetting events.  In model A we determine  exact distributions of these two quantities.  The variance of the center of mass for both models scales as $1/N$. The variance of the radius is independent of $N$ in model A and exhibits an unusual scaling $(\ln N)/N$ in model B. The latter scaling is intimately related to the $1/f$ noise in the radius autocorrelations.
Finally, we evaluate the mean first-passage time (MFPT) to a distant target in model A, model B, and the Brownian bees model. For model A we obtain an exact asymptotic expression for the MFPT which scales as $1/N$. For model B and the Brownian bees model we propose a sharp upper bound for the MFPT. The bound assumes an ``evaporation" scenario, where the first passage requires
multiple attempts of a single particle, which breaks away from the rest of the particles, to reach the target. The resulting MFPT for model B and the Brownian bees model scales exponentially with $\sqrt{N}$.
We verify this bound by performing highly efficient weighted-ensemble simulations of the first passage in model B.
\end{abstract}

\maketitle

\section{Introduction} \label{sec:intro}

Physicists have always strived for universality, and this quest is fully present in statistical mechanics of many-particle systems out of equilibrium. In this work we propose a unified approach to two different families of many-particle models in space which have attracted much interest in recent years.  The first family of models deals with $N$ independent Brownian particles which undergo stochastic resetting  to a specified point in space ~\cite{evans2011diffusion,evans2011diffusion2,bhat2016stochastic,evans2020stochastic}. The original motivation behind these models was to optimize random search of a target. Indeed,  the mean first passage time (MFPT) to a stationary target is infinite without reset \cite{redner2001guide}, but it becomes finite once reset is introduced \cite{evans2011diffusion,evans2011diffusion2,bhat2016stochastic,evans2020stochastic}.  Apart from the random search optimization, the reset models provide an interesting example of the emergence of a nonequilibrium steady state (NESS)~\cite{krapivsky2010kinetic, mendez2016characterization}, and this feature will play a prominent role in most of our paper.

The second, and seemingly unrelated, family of models deals with branching Brownian motion (BBM) of $N$ particles with selection \cite{mckean1975application, bramson1983convergence, brunet2009statistics}, where in each branching event, the particle with the lowest fitness is removed, so that the total number of particles remains constant \cite{brunet2006noisy, brunet2007effect, berard2010brunet, durrett2011brunet}. The members of this family of models differ from each other by the choice of fitness function, which mimics different aspects of biological selection. A recent example of a simple, yet nontrivial, model of this class is the ``Brownian bees" \cite{berestycki2020brownian, berestycki2020free, siboni2021fluctuations, meerson2021persistent}. In this model, when a branching  event occurs, the particle which is farthest from the origin is removed.

A close connection between the two families of models becomes obvious upon observation that, for example, the Brownian bees model can be easily reformulated as a reset model. Indeed, the combined process of a branching event and the removal of the farthest particle is equivalent to resetting the farthest particle to the exact location of any of the remaining $N-1$ particles.

Here we study  stationary fluctuations in two models. The first is the ``classical" model (denoted as model A) where the Brownian particles reset to the origin independently \cite{evans2011diffusion,evans2011diffusion2}. The second is
a new model (model B) that we introduce here, which is a relative of both model A and the Brownian bees model. As in the latter, only the particle farthest from the origin can undergo reset. Yet particles are reset only to the origin, as in model A. The selection of only the farthest particles as candidates for reset -- both in model B, and in the Brownian bees model -- introduces nonlocal correlations between the particles. This is in contrast to model A, where the particles are reset independently from each other.

At long times models A and B approach their NESSs. A crucial difference between these models appears
already in the hydrodynamic limit, $N\to \infty$. Here the steady-state coarse-grained particle density in model A has an infinite support, while that in model B has a compact support similarly to the Brownian bees model \cite{berestycki2020brownian, berestycki2020free, siboni2021fluctuations}. In model A, a finite system radius, which scales as $\ln N$, appears only when $N$ is finite.   In models A and B we study stationary fluctuations of the center of mass of the system, and of the system's radius due to the random character of the Brownian motion and of the resetting events. For model A we determine exact distributions of these two quantities. We show that, as in the Brownian bees model \cite{siboni2021fluctuations}, the variance of the center of mass for models A and B scales as $1/N$. The variance of the radius in model A is independent of $N$, whereas in model B it exhibits the same anomalous scaling $(\ln N)/N$ as the variance of the radius in the Brownian bees model. This anomalous scaling
is intimately related to the $1/f$ noise in the radius autocorrelations, and it originates from the compact support
of the hydrodynamic steady state solution. Strikingly, the numerical coefficients of the autocorrelation functions of the radius in model B and in the Brownian bees model coincide, which suggests a certain universality of the distribution of the radius in this class of models.

Finally, we return to the original motivation behind the reset models and study the MFPT to a \emph{distant} target, at $x=L$, in models A and B and the Brownian bees model. For model A we employ known single particle results \cite{evans2011diffusion} to obtain an exact asymptotic expression for the MFPT, $\langle T\rangle \simeq (1/N)\,e^L$, where $\langle T\rangle$ is assumed to be large.   For model B and the Brownian bees model the MFPT is much longer. We propose a sharp upper bound for it, determined by repeated attempts of a single breakaway particle to reach the target. The resulting MFPT scales exponentially with $\sqrt{N}$, rather than with $N$. Therefore, the single-particle evaporation scenario is exponentially more efficient than any macroscopic scenario which involves ${\cal O}(N)$ effective number of particles. We verify this bound by performing highly efficient weighted-ensemble simulations of the first passage in model B.

We obtain our analytical results by using exact probabilistic calculations (for model A) and a coarse-grained Langevin-type description (for both models). The coarse-grained description is expected to hold at length scales much larger than the typical interparticle distance, and on time scales much longer than the inverse reset rate.  In this sense the exact ``microscopic" calculations for model A also provide a good benchmark for the approximate coarse-grained description.

Here is a layout of the remainder of the paper. Steady-state fluctuations in models A and B are dealt with in Secs. \ref{sec:independent} and \ref{sec:InteractingParticles}, respectively. In Sec. \ref{MFPT} we study the MFPT to a distant target for models A and B and for the Brownian bees model. We summarize our results and discuss some unresolved issues in Sec.  \ref{discussion}.

\section{Model A: independent resets} \label{sec:independent}
\begin{figure}[t!]
\centering
\includegraphics[width=.80\linewidth]{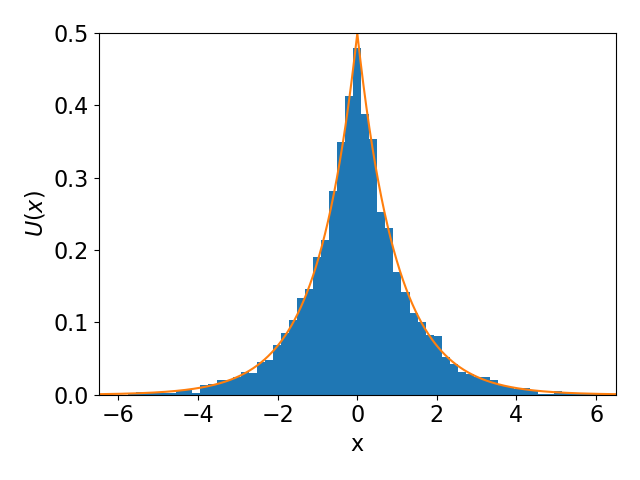}
\vspace{-3mm} \caption{A Monte Carlo realization of $N= 10^4$ non-interacting resetting Brownian particles (model A, blue bars) is compared with the deterministic steady state solution \eqref{Eq3} (solid line) for $r = D = 1$. The simulation snapshot is taken at $t=1000$. In this and all other MC simulations, reported in this work,  all the particles start at $t=0$ from $x=0$.}
 \label{fig:fig1}
\end{figure}

We start with the well-known model \cite{evans2011diffusion} of $N$ independent Brownian particles on a line, with a diffusion constant $D$,
each undergoing resetting to the origin $x = 0$ at rate $r$. This reset is equivalent to two effective elemental processes, perfectly synchronized in time: independent death of a particle in the bulk,
and a simultaneous arrival of a new particle to the origin. For  $N\to \infty$, the particles' coarse-grained spatial density $u(x,t)$  in this model \footnote{The coarse-grained density $u(x,t)$ describes the system on length scales much larger than the typical inter-particle distance.} is governed by the  continuous deterministic equation
\begin{eqnarray}
    &\pp_t u(x, t) = D\pp^2_x u(x, t) - r u(x, t) + r N \delta(x). \label{Eq1original}
\end{eqnarray}
Exactly the same equation describes the evolution of the probability distribution of the position of a single Brownian particle subject to reset to the origin \cite{evans2011diffusion}. This is a natural consequence of the particle independence in model A. As to be expected, Eq.~\eqref{Eq1original} obeys the conservation law
\begin{equation}
 \int_{-\infty}^{\infty} u(x, t) dx = N. \label{Eq2original}
\end{equation}
For convenience, we recast Eqs.~(\ref{Eq1original}) and (\ref{Eq2original}) in a dimensionless and normalized form by defining an inverse length scale as $\kappa = \sqrt{r/D}$, and the rescaled variables $\tilde{x} = \kappa x$, $\tilde{t} = rt$ and $\tilde{u} = u \kappa^{-1}/N$. In these variables Eqs.~(\ref{Eq1original}) and (\ref{Eq2original}) become
\begin{eqnarray}
    &\pp_{\tilde{t}} \tilde{u}(\tilde{x}, \tilde{t}) = \pp^2_{\tilde{x}} \tilde{u}(\tilde{x}, \tilde{t}) - \tilde{u}(\tilde{x}, \tilde{t}) + \delta(\tilde{x}) ,\label{Eq1}\\
    &\int_{-\infty}^{\infty} \tilde{u}(\tilde{x}, \tilde{t}) d\tilde{x} = 1 . \label{Eq2}
\end{eqnarray}
At long times the solution to Eqs. (\ref{Eq1}) and (\ref{Eq2}) approaches a unique steady state density given by
\begin{equation} \label{Eq3}
    \tilde{U}(\tilde{x}) = \frac{1}{2} e^{-|\tilde{x}|}
    \Rightarrow
    U(x) = \frac{1}{2} N \kappa e^{-\kappa|x|}    ,
\end{equation}
which is nothing but the sum over $N$ independent particles  of the NESS of a single particle \cite{evans2011diffusion}. Figure \ref{fig:fig1} compares Eq.~\eqref{Eq3} with a long-time snapshot of a Monte Carlo simulation of the microscopic model.
In the following, unless otherwise specified, we omit all tildes for brevity.

Now we consider steady-state fluctuations of this system at  large but finite $N$. One of the quantities of our interest is the center of mass $X(t)$. Because of the reflection symmetry $x\to -x$ of the microscopic model,  the average value of $X(t)$ is zero. Therefore, we focus on the two-time autocorrelation
of $X(t)$, defined as $ g_X(t_1, t_2) = \average{X(t_1)X(t_2)}$. For $t_1, t_2 \gg 1$, \textit{i.e.}, times much longer than the typical relaxation time, the NESS is reached, and the autocorrelation depends only on the time difference $\tau=t_1-t_2 $:
\begin{equation} \label{corr_def}
    g_X(\tau) = \average{X(0)X(\tau)}.
\end{equation}

To address typical, small fluctuations of $X(t)$, one can continue using a coarse-grained description of the system
in terms of the particle density $u(x,t)$, which now becomes a stochastic field.  This field is governed by the (rescaled) Langevin equation
\begin{equation} \label{Eq4}
    \pp_t u(x, t) = \pp^2_x u(x, t) - u(x, t) + \delta(x) + R_I(u, x, t) ,
\end{equation}
which replaces Eq.~\eqref{Eq1}.  The noise term $R_I=R_{BM}+R_{d} + R_{a} $ includes three contributions:
\begin{eqnarray}
    &&R_{BM} = \frac{1}{\sqrt{N}} \pp_x \left( \sqrt{2 u}\, \chi(x, t) \right), \label{Rf}\\
    &&R_{d} = \frac{\sqrt{u}}{\sqrt{N}}
    \,\eta(x, t), \label{Rd} \\
    &&R_{a} = -\frac{\delta(x)}{\sqrt{N}} \int_{-\infty}^{\infty}\!\! \sqrt{u(x',t)}\, \eta(x', t)\,dx',\label{Rm}
\end{eqnarray}
where $\chi(x, t)$ and $\eta(x, t)$ are two independent Gaussian white noises with zero mean:
\begin{eqnarray}
    \average{\chi(x_1, t_1)\chi(x_2, t_2)} &=& \average{\eta(x_1, t_1)\eta(x_2, t_2)} \nonumber \\
    &=& \delta(x_1 - x_2)\delta(t_1 - t_2).
\end{eqnarray}
The term $R_{BM}$ is the Brownian motion noise term. It is best known in the context of a large-scale and long-time description of the
stochastic dynamics of a lattice gas of independent random walkers \cite{spohn2012large,meersonsasorov2011}.
The term $R_d$
originates from the exact master equation for the death process. The derivation procedure (see \textit{e.g.,} Refs.~\cite{gardiner1985handbook,assaf2017wkb}) starts from applying  the van Kampen system-size expansion to approximate (for typical fluctuations and $N \gg 1$)  the exact discrete master equation by a continuous Fokker-Planck equation. The latter is equivalent to the Langevin description with noise term $R_d$.
Finally, the term $R_{a}$ -- the arrival noise -- describes the noise in the particle arrival at $x=0$. Its magnitude is determined by the demand that the total number of particles be conserved at any time in the presence of the death process.

\begin{figure}[t!]
\centering
\includegraphics[width=1.03\linewidth]{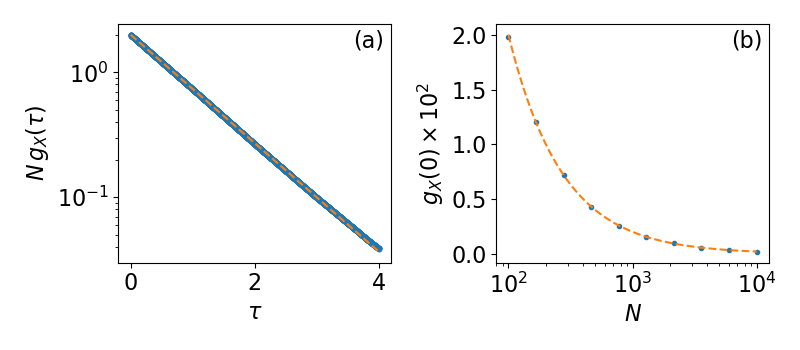}
 \vspace{-5mm}\caption{The steady-state autocorrelation function of the center of mass of a system of non-interacting Brownian particles under reset (model A). The simulation results (points) are plotted as a function of the time delay $\tau$ for $N = 10^4$  (a) and  as a function of $N$ for $\tau = 0$ (b). The dashed lines show the predictions of Eq.~\eqref{corr_nocorr}. Note the logarithmic scale in (a). }
 \label{fig:fig2}
\end{figure}

The Langevin equation~(\ref{Eq4}) is a nonlinear stochastic integro-differential equation in partial derivatives for $u(x,t)$. However, it simplifies dramatically if we exploit the small parameter $1/\sqrt{N} \ll 1$ in the noise terms (\ref{Rf})-(\ref{Rm}) and perform a perturbation expansion of $u(x,t)$ around the deterministic steady state $U(x)$. Setting
\begin{equation}
    u(x, t) = U(x) + v(x, t), \; \; |v| \ll 1
\end{equation}
and linearizing Eq.~\eqref{Eq4},
leads to a linear stochastic partial differential equation:
\begin{equation} \label{linearizedEquation}
    \pp_t v(x, t) = \pp^2_x v(x, t) - v(x, t) + R_I( x, t),
\end{equation}
where we have denoted $R_I(x, t) = R_I(U(x), x, t)$.  Being interested in steady-state quantities, we can set the initial conditions as $v(x, t_0) = 0$ and send $t_0\to -\infty$. The resulting solution for $v(x, t)$, for given realizations of the noises, can be written as
\begin{equation}
    v(x,t) = \int_0^\infty \int_{-\infty}^{\infty} \frac{1}{\sqrt{4 \pi t'}} \,e^{-t'- \frac{(x-x')^2}{4t'}}R_I(x', t-t') dx' dt'.
    \label{vmodelA}
\end{equation}
The center of mass $X(t)  = \int_{-\infty}^\infty x u(x,t) dx$
can be determined as follows:
\begin{equation} \label{COMnocorrelation0}
    X(t)  = \int_{-\infty}^\infty x \left[U(x,t) +v(x,t)\right]dx = \int_{-\infty}^\infty x v(x,t) dx .
\end{equation}
Plugging here Eq.~(\ref{vmodelA}) and performing the integration over $x$, we obtain
\begin{equation} \label{COMnocorrelation}
    X(t) = \int_0^\infty \int_{-\infty}^{\infty} x'e^{-t'}R_I(x', t-t') dx' dt'.
\end{equation}
Now we plug Eq.~\eqref{COMnocorrelation} into  Eq.~\eqref{corr_def} and perform the averaging over the noise. The terms proportional to $\delta(x)$ do not contribute, and  after some algebra we arrive at a simple result:
\begin{equation} \label{corr_nocorr}
    g_X(\tau) = \frac{2}{N}e^{-\tau}.
\end{equation}
Figure \ref{fig:fig2} shows that Eq.~\eqref{corr_nocorr} agrees very well with MC simulations for all $\tau$.
The variance of the center of mass is given by $\average{X^2} = g_X(0) =  2/N$. The $1/N$ scaling of the variance of the center of mass in model A is to be expected
from the law of large numbers. The same scaling is observed for the Brownian bees \cite{siboni2021fluctuations} and, as we show below, for model B.

Since the particles in model A are independent,  Eq.~(\ref{corr_nocorr}) can be also obtained from the autocorrelation function for a single particle that performs Brownian motion and is reset to the origin, as in the original reset model \cite{evans2011diffusion}. The single-particle autocorrelation function was calculated in Ref. \cite{MajumdarOshanin}, and our result (\ref{corr_nocorr}) perfectly agrees with their calculations. As we see,  linearization of the Langevin equation yields an exact result in this case.

The exceptionally simple expression (\ref{corr_nocorr}) for the  autocorrelation hints at a possible interpretation of fluctuations of the center of mass of the system in terms of an effective Ornstein-Uhlenbeck process. Indeed,
multiplying both sides of Eq.~\eqref{linearizedEquation} by $x$ and integrating over $x$, we obtain
\begin{equation}
  \dot{X}(t) = -X(t)
 +\frac{1}{\sqrt{N}}\!\int_{-\infty}^{\infty} \!\!\!dx\,\sqrt{U(x)}\left[\sqrt{2} \chi(x,t)\!+\! x \eta(x,t)\right]\!.
  \label{OU}
\end{equation}
This is the Ornstein-Uhlenbeck equation, and the center of mass effectively behaves as a single ``macroparticle" undergoing an overdamped motion in a quadratic potential $X^2/2$  under  properly weighted Gaussian white noises $\chi(x,t)$ and $\eta(x,t)$.

The particles' independence in model A also makes it possible to calculate
\emph{exact} steady-state probability distribution of the center of mass. Indeed, the joint probability distribution $P(x_1, x_2, \dots, x_N)$ of the positions of $N$ independent resetting particles in the steady state immediately follows from the single-particle distribution,
\begin{equation}\label{singleprob}
P(x_1, x_2, \dots, x_N)= 2^{-N}  \exp\left(-\sum_{i=1}^{N} |x_i|\right)\, ,
\end{equation}
[see Eq.~(\ref{Eq3})]. We are interested in the probability distribution $\mathcal{P}(a,N)$ that the observed position of the center of mass,
\begin{equation}\label{comcondition}
X \equiv \frac{1}{N}\sum_{i=1}^{N} x_i\,,
\end{equation}
is equal to a specified value $a$. This distribution is given by the integral
\begin{eqnarray}
\mathcal{P}(a,N)  &=& 2^{-N}\,\int_{-\infty}^{\infty} dx_1 \int_{-\infty}^{\infty} dx_2\,
\dots \int_{-\infty}^{\infty} dx_N  \nonumber\\
 &\times& \exp\left(-\sum_{i=1}^{N} |x_i|\right)\,\delta \left(\frac{1}{N}\sum_{i=1}^N x_i -a\right). \label{proba}
\end{eqnarray}
To evaluate this integral, we use the exponential representation of the delta-function,
$\delta(z)=(2\pi )^{-1}\int_{-\infty}^{\infty} e^{ikz} \,dz$, leading to
\begin{eqnarray}
  \mathcal{P}(a,N)  &=& \frac{2^{-N} N}{2\pi}\, \int_{-\infty}^\infty dk \,e^{-ikNa} \nonumber \\
  &&\times \prod_{n=1}^{N} \int_{-\infty}^{\infty} dx_n \exp\left(- |x_n| +   i k x_n \right)\,.\label{proba1}
\end{eqnarray}
The factorized integrals over $x_n$ are elementary, and the calculation reduces to a single integration over $k$:
\begin{equation}\label{proba3}
  \mathcal{P}(a,N) =\frac{N}{2\pi}\int_{-\infty}^\infty dk \,
  e^{-N\left[i k a- \ln(1+k^2)\right]}\,.
\end{equation}
This integration can be performed exactly, and we obtain the exact distribution,
\begin{equation}\label{proba4}
 \mathcal{P}(a,N) =  \frac{2^{\frac{1}{2}-N} N^{N+\frac{1}{2}} | a| ^{N-\frac{1}{2}}
   K_{N-\frac{1}{2}}(N | a| )}{\sqrt{\pi} \,(N-1)!}\,,
\end{equation}
valid for any number of particles. Here $K_{\alpha}(z)$ is the modified Bessel function with argument $z$ and index $\alpha$. As $\alpha =N-1/2$ is half-integer, $K_{\alpha}(z)$ can be expressed as a finite sum of elementary functions.

The $N\gg 1$ limit of $\mathcal{P}(a,N)$ can be probed by extracting the $N\gg 1$ asymptotic of Eq. (\ref{proba4}). A more aesthetically pleasing alternative, however, is to exploit the large parameter $N$ directly in Eq.~(\ref{proba3}) and use the saddle point method in the complex plane. The relevant saddle point is
\begin{equation}\label{saddlepoint}
k=k_*(a)= i \left(\frac{1-\sqrt{a^2+1}}{a}\right)\,.
\end{equation}
Ignoring the pre-exponential factors in Eq.~(\ref{proba3}), this leads to
\begin{equation}\label{LDscaling}
-\ln  \mathcal{P}(a,N\gg 1)  \simeq N \Phi(a)\,,
\end{equation}
with the rate function  \footnote{The same expression for $\Phi(a)$ was also obtained by Satya N. Majumdar (private communication).}
\begin{equation}\label{LDF}
\Phi(a)  = \sqrt{a^2+1}+\ln \left[\frac{2
   \left(\sqrt{a^2+1}-1\right)}{a^2}\right]-1\,.
\end{equation}
As to be expected, the rate function vanishes at $a=0$. Further, it is quadratic at small $a$, $\phi(a\to 0) \simeq a^2/4$. This corresponds to the Gaussian part of the distribution of the center of mass with the variance $2/N$ in agreement with the macroscopic result~(\ref{corr_nocorr}). In fact, this result immediately follows from law of large numbers and the known variance (equal to $2$) of the single-particle distribution of $x$.  At large $a$ we obtain  $\Phi(a) \simeq |a|$, which describes exponential large-deviation tails of the distribution.

Let us now briefly return to the deterministic continuous description, see Eq.~(\ref{Eq1original}).
The steady-state density distribution, predicted by Eq.~(\ref{Eq3}), lives on the whole line $|x|<\infty$. Yet, the actual system radius $\ell(t)$ in the original microscopic model --  the absolute value of the position of the particle farthest from the origin -- is of course finite at all times, and this effect is missed by the continuous model (\ref{Eq1original}).
Still, the average system's radius in the steady state, $\average{\ell}$, can be readily found, with logarithmic accuracy, from the equation $\int_{\langle{\ell}\rangle} ^{\infty} U(x) \,dx = 1/N$, with
$U(x)$ given by the continuum equation~(\ref{Eq3}). The result is $\average{\ell} \simeq \ln N$.

\begin{figure}[t]
\includegraphics[width=0.4\textwidth,clip=]{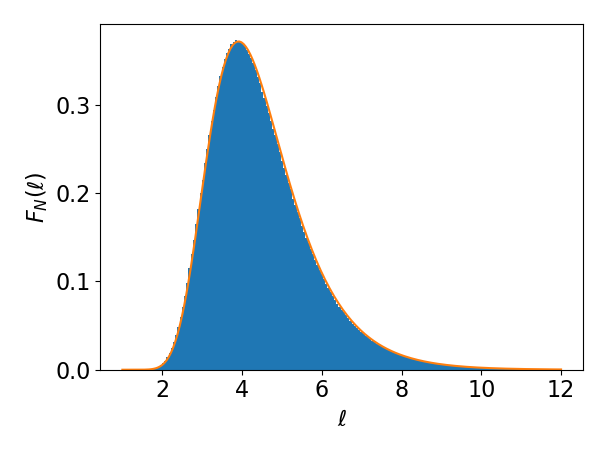}
\caption{The distribution of the system's radius in model A: simulations (blue bars) are compared with $F_N(\ell)$, Eq.~(\ref{FN}) (solid line) for $N=50$.}
\label{Fplot}
\end{figure}

In fact, one can not only improve this leading-order estimate of $\average{\ell}$, but also calculate
\emph{exact} steady-state probability distribution of $\ell$ in model A. Indeed, using the probability $p(x) = (1/2)\,e^{-|x|}$ of the position of a single resetting particle in the steady state, we can easily calculate
the \emph{cumulative} probability to observe the particle within a specified range  $|x|<\ell$:
\begin{equation}\label{cum1}
Q_1(\ell) = \int_{-\ell}^{\ell}  dx \,p(x) = 1-e^{-\ell}\,.
\end{equation}
As the $N$ particles are independent, the multi-particle cumulative probability is
\begin{equation}\label{cumN}
Q_N(\ell) = Q_1(\ell)^N = \left(1-e^{-\ell}\right)^N\,.
\end{equation}
The exact probability density of $\ell$ is, therefore
\begin{equation}\label{FN}
F_N(\ell) = \frac{dQ_N(\ell)}{d\ell}=N e^{-\ell} \left(1-e^{-\ell}\right)^{N-1}\,.
\end{equation}
Figure \ref{Fplot} shows an example of this distribution for $N=50$, compared with MC simulations.

The first moment of this distribution (the mean value of $\ell$) is
\begin{equation}\label{av}
\average{\ell}= \int_0^{\infty} d\ell\, \ell \, F_N(\ell) = H_N,
\end{equation}
the $N$-th harmonic number. For $N\gg 1$ we obtain
\begin{equation}\label{av1}
\average{\ell} = \ln N+\gamma+ O\left(\frac{1}{N}\right),
\end{equation}
where $\gamma=0.57721\dots$ is the Euler's constant.

The second moment of the distribution $F_N(\ell)$ is
\begin{equation}\label{second}
\average{\ell^2}= \int_0^{\infty} d\ell\, \ell^2 \, F_N(\ell) = \left(H_N\right){}^2-\psi ^{(1)}(N+1)+\frac{\pi ^2}{6}\,,
\end{equation}
where $\psi^{(1)}(\dots)$ is the polygamma function. At large $N$
\begin{equation}\label{second1}
\average{\ell^2} = \ln^2N+2 \gamma  \ln N+\frac{\pi ^2}{6}+\gamma ^2+O\left(\frac{\ln N}{N}\right)\,.
\end{equation}
The variance of $\ell$,
\begin{equation}\label{var1}
\text{Var}(\ell) \equiv \average{\ell^2}- \average{\ell}^2 = \frac{\pi^2}{6}-\psi ^{(1)}(N+1)\,,
\end{equation}
behaves at large $N$ as
\begin{equation}\label{var2}
\text{Var}(\ell) = \frac{\pi^2}{6}+O\left(\frac{1}{N}\right)\,.
\end{equation}
The leading-order term  $\pi^2/6$ coincides with the variance of the standard Gumbel distribution. This is not surprising in view of the fact that the single-particle distribution in model A (the ``parent distribution") decays exponentially with $x$ \cite{Majumdarreview}.

\begin{figure}[t!]
\centering
\includegraphics[width=1.0\linewidth]{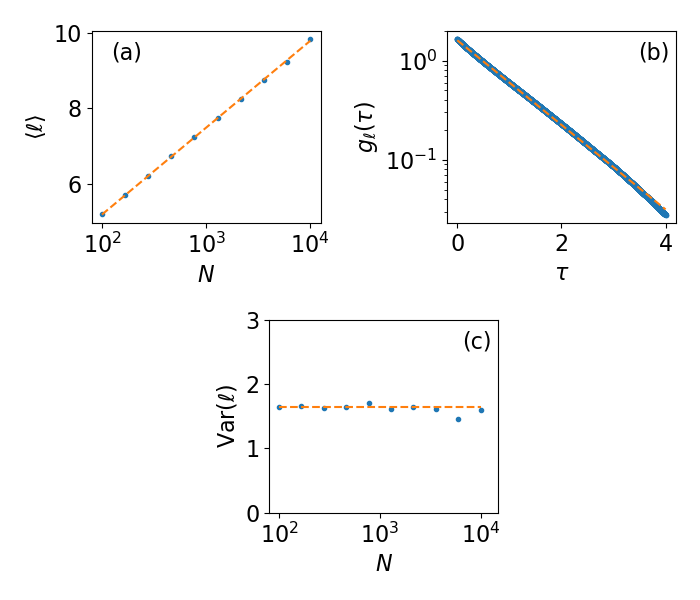}
 \vspace{-7mm}
 \caption{Non-interacting Brownian particles under reset (model A). (a) The average system's radius $\langle \ell \rangle$ in the steady state. The simulation results (points) are plotted as a function of $N$ on the logarithmic scale. The dashed line shows the theoretical prediction (\ref{av1}). (b) The autocorrelation function of the system's radius  as a function of $\tau$ for $N = 10^4$ (logarithmic scale), fitted by a single exponential function (dashed line, see text). (c)  The variance of $\ell$ as measured in the simulations (points) versus the prediction $\pi^2/6$, see Eq.~(\ref{var2}) (dashed line).}
 \label{fig:radiusA}
\end{figure}

As to be expected, the predictions (\ref{av1}) and (\ref{var2}) agree very well with our MC simulations of model A, see  Fig.~\ref{fig:radiusA}, panels (a) and (c).  The steady-state autocorrelation function of $\ell$, $g_{\ell}(\tau)$ appears to display a purely exponential decay with the exponent close to $1$: $g_{\ell}(\tau) \simeq 1.61  \exp(-0.99\tau)$, see Fig.~\ref{fig:radiusA}(b).   The fitted coefficient $1.61$ is close to the theoretically predicted variance  $\text{Var}(\ell)  \simeq \pi^2/6 \simeq 1.64$.


\section{Model B: reset of farthest particle} \label{sec:InteractingParticles}
We now consider model B. To remind the reader, here at each random resetting event the particle farthest from the origin is
reset to zero. In the limit of $N\to \infty$, the rescaled and normalized coarse-grained spatial particle density $u(x,t)$ is governed by the deterministic free-boundary problem
\begin{eqnarray}
    &\pp_t u(x, t) = \pp^2_x u(x, t) + \delta(x), \quad  \;\; |x| \leq \ell(t),  \label{Eq1interacting}\\
    & u(x, t) = 0, \quad  \;\; |x| >\ell(t), \label{Eq2interacting}\\
    &\int_{-\ell(t)}^{\ell(t)} u(x, t) dx = 1, \label{Eq3interacting}
\end{eqnarray}
where $u(x,t)$ is continuous at $x = \ell(t)$~\footnote{Note that in this rescaled form $\ell$ should actually be $\tilde{\ell}$ defined by $\tilde{\ell} = \kappa^{-1}\ell$, and similarly for all other quantities. We however omit all tildes unless otherwise specified.}. As in the case of the Brownian bees model \cite{berestycki2020brownian, siboni2021fluctuations}, here too the coarse-grained density $u(x,t)$ lives on the compact support $|x|<\ell(t)$.  The effective absorbing walls at $x=\pm \ell(t)$ move in synchrony to maintain a constant number of particles. The emergence, in the hydrodynamic limit, of compact support is a direct consequence of the reset of the farthest particle.

\begin{figure}[t!]
\centering
\includegraphics[width=.80\linewidth]{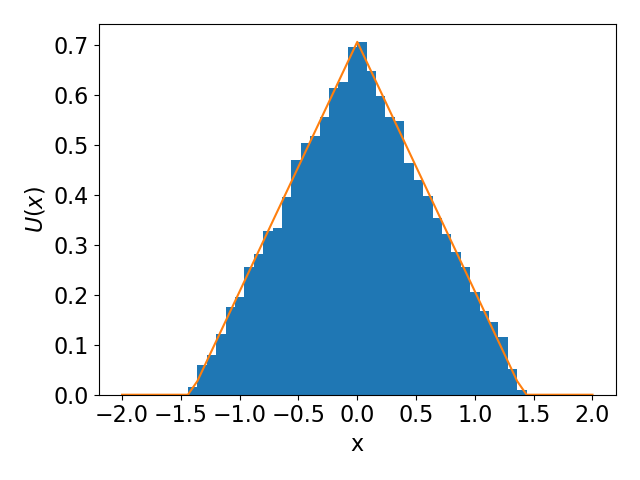}
 \vspace{-3mm}\caption{A Monte Carlo realization of $N= 10^4$ interacting Brownian particles (model B, blue bars) is compared with the deterministic steady state solution  \eqref{steady_state_ineracting} (solid line) for $r = D = 1$. The simulation snapshot is taken at $t=100$.}
 \label{fig:fig3a}
\end{figure}

At long times, the solution of the problem (\ref{Eq1interacting})-(\ref{Eq3interacting}) approaches a unique steady state
\begin{numcases}
{{U(x)} =}\frac{1}{2}
     \left(\ell_0 - |x|\right), & $|x| \le \ell_0$, \label{steady_state_ineracting} \\
0\,,& $|x|>\ell_0$, \label{outside}
\end{numcases}
whereas $\ell(t)$ approaches $\ell_0$. Here $\ell_0= \sqrt{2}$ is the (rescaled) radius of the system in the limit of $N\to \infty$.
In Fig. \ref{fig:fig3a} we compare Eqs.~(\ref{steady_state_ineracting})-(\ref{outside}) with a late-time snapshot from MC simulations.

Because of the inter-particle correlations in model B, exact probabilistic calculations are hardly possible. Therefore we assume that $N$ is large (but finite) and use the coarse-grained description of the system in terms of a Langevin equation. The Langevin equation to replace Eq.~\eqref{Eq1interacting} is
\begin{equation} \label{EqUinteracting}
    \pp_t u(x, t) = \pp^2_x u(x, t) + \delta(x) + R(u, x, t),\quad |x| \leq \ell(t),
\end{equation}
while Eqs. (\ref{Eq2interacting}) and (\ref{Eq3interacting}) remain unchanged.   The noise term $R = R_{BM} + R_{a}$ originates from two independent noises given by Eqs. \eqref{Rf} and \eqref{Rm}, as discussed in Sec. \ref{sec:independent}. The calculation procedure below closely follows that used in Ref. \cite{siboni2021fluctuations} for the Brownian bees model.  For convenience we differentiate the conservation law \eqref{Eq3interacting} with respect to time, and use Eq.~\eqref{EqUinteracting} to arrive at the equation
\begin{equation} \label{Eq3diff}
    \pp_x u[-\ell(t), t] - \pp_x u[\ell(t), t] = 1 + \int_{-\ell(t)}^{\ell(t)} R(u, x, t) \;dx,
\end{equation}
which
replaces Eq.~\eqref{Eq3interacting}.

Employing the small parameter $1/\sqrt{N} \ll 1$ and linearizing Eq.~\eqref{EqUinteracting} around the steady state \eqref{steady_state_ineracting}, we find:
\begin{eqnarray}
    u(x, t) = U(x) + v(x, t), \quad\quad |v| \ll 1,\\
    \ell(t) = \ell_0 + \delta \ell(t)     , \quad\quad  |\delta \ell(t)| \ll 1.
\end{eqnarray}
Plugging these into  Eqs.  \eqref{Eq2interacting}, \eqref{EqUinteracting} and \eqref{Eq3diff}, we obtain the following linearized equations
\begin{eqnarray}
    &\pp_t v(x, t) - \pp^2_x v(x, t) = R(x, t),  \label{linearized_1} \\
   & v(\pm \ell_0, t) = - \frac{1}{2}\delta \ell(t), \label{linearized_2} \\
   & \pp_x v(-\ell_0, t) - \pp_x v(\ell_0, t) = \int_{-\ell_0}^{\ell_0} R(x, t) \;dx, \label{linearized_3}
\end{eqnarray}
where $R(x, t) = R(U(x), x, t)$. Again,  at long times we can set the initial condition as $v(x, t_0) = 0$ and send $t_0 \to -\infty$. We rewrite the conditions in Eq.~\eqref{linearized_2} as
\begin{equation} \label{linearized_2_fixed}
    v(-\ell_0, t) = v(\ell_0, t)  \quad \text{and} \quad \delta \ell(t) = 2  v(\ell_0, t).
\end{equation}
The first of these relations allows one to continuously extend $v(x,t)$ and $R(x,t)$ to the whole $x$ axis periodically. While $v$ is continuous, its $x$-derivative has finite jumps at $x = \ell_0(1+2m)$ for any $m = 0, \pm 1, ...$, see Eq.~\eqref{linearized_3}. To account for these jumps we need to modify the source term in Eq.~\eqref{linearized_1}. We define $\tilde{R} = \tilde{R}_{BM} + \tilde{R}_a$ with
\begin{equation}
    \tilde{R}_i = R_i(x,t) - \left(\int_{\ell_0}^{\ell_0}R_i(x,t) dx\right) \sum_{m\in\mathcal{Z}} \delta\left(x - \ell_0(1+2m)\right)
\end{equation}
for $i = \{BM, a\}$, such that Eq.~\eqref{linearized_1} becomes
\begin{equation}
    \pp_t v(x, t) - \pp^2_x v(x, t) = \tilde{R}(x, t).  \label{linearized_1_fixed}
\end{equation}
Note that the new source term $\tilde{R}$ obeys the equation $\int_{-\ell_0+\Delta}^{\ell_0 + \Delta}\tilde{R}(x, t)dx = 0$ for any real $\Delta$.
We can then further simplify the problem by shifting the interval of interest $[-\ell_0, \ell_0]$ by an infinitesimal $\Delta$ such that Eq.~\eqref{linearized_3} becomes
\begin{equation} \label{linearized_3_fixed}
    \pp_x v(-\ell_0 + \Delta, t) - \pp_x v(\ell_0 + \Delta, t) = 0 ,
\end{equation}
for $\Delta \to 0$.

\begin{figure}[t!]
\centering
\includegraphics[width=1.03\linewidth]{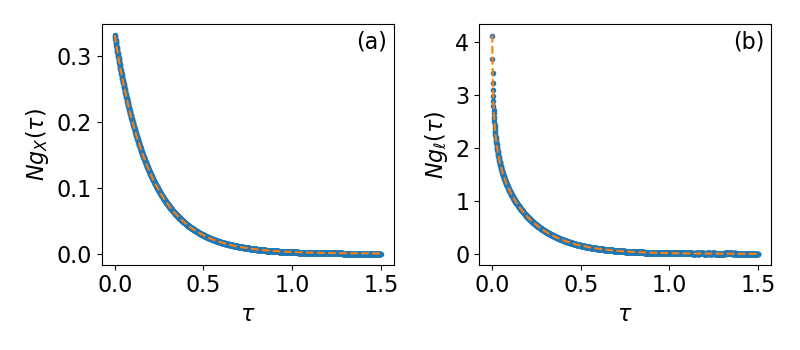}
 \vspace{-7mm}\caption{The autocorrelation functions in model B of (a) the system's center of mass and (b) the system's radius (simulations, points) for $N = 10^4$, compared with Eq.~\eqref{COMswarm_result} in (a) and Eq.~\eqref{CovEll} in (b)  (dashed lines). }
 \label{fig:fig3}
\end{figure}

The problem  defined by Eqs. \eqref{linearized_1_fixed}, \eqref{linearized_2_fixed} and \eqref{linearized_3_fixed} can be solved using the Green's function formalism~\cite{siboni2021fluctuations}. We expand over the eigenfunctions of the linear operator in Eq.~\eqref{linearized_1_fixed}, given by $\{1, \cos(\pi n x/ \sqrt{2}), \sin(\pi n x/ \sqrt{2})\}$, with corresponding eigenvalues  $\lambda_n = \pi^2 n^2/2$, for $n \in \mathbf{N}$.
The solutions for $v$ and $\delta \ell$ are
\begin{eqnarray} \label{v_interacting}
    v(x,t) = \frac{1}{\sqrt{2}} \sum_{n=1}^\infty \int_0^\infty \int_{-\sqrt{2}}^{\sqrt{2}} &&e^{-\lambda_n t'} \tilde{R}(x', t-t')  \\ && \times \cos[\sqrt{\lambda_n} (x - x')] dx' dt', \nonumber
\end{eqnarray}
\begin{eqnarray} \label{ell_interacting}
    \delta \ell(t) = \sqrt{2} \sum_{n=1}^\infty (-1)^n \int_0^\infty \int_{-\sqrt{2}}^{\sqrt{2}} &&e^{-\lambda_n t'} \tilde{R}(x', t-t') \\ && \times \cos(\sqrt{\lambda_n} x') dx' dt'.\nonumber
\end{eqnarray}
We now use these results to determine the autocorrelation functions for the system's center of mass and radius. Plugging  Eq.~\eqref{v_interacting} into Eq.~(\ref{COMnocorrelation0}), we express the center of mass $X(t)$ as
\begin{eqnarray}
    &&X(t) = \frac{2 \sqrt{2}}{\pi } \sum _{n=1}^{\infty } \frac{(-1)^{n+1}}{n} f_n(t),\\
    &&f_n(t) = \int _{-\sqrt{2}}^{\sqrt{2}}\int
   _0^{\infty }e^{-\lambda_n t'} \sin \left(\sqrt{\lambda_n} x\right)
   \tilde{R}\left(x,t-t'\right)dt' dx. \nonumber
\end{eqnarray}
Plugging this expression into Eq.~\eqref{corr_def}, averaging over the noise, and summing up one of the two sums, we find
\begin{equation} \label{COMswarm_result}
    g_X(\tau) = \frac{32}{\pi^4 N} \sum_{n= 0}^\infty \frac{1}{(2n+1)^4}e^{-\frac{\pi ^2}{2}  (2n+1)^2 \tau }.
\end{equation}
This autocorrelation is qualitatively similar to that for the Brownian bees \cite{siboni2021fluctuations}. It is more complicated than that for model A, see Eq.~(\ref{corr_nocorr}),  because of the discrete  spectrum of the linearized operator of model B, in contrast
to the continuous spectrum of model A.  In particular, $X(t)$ of model B behaves as an Ornstein-Uhlenbeck ``macroparticle" only in the limit of $\tau\gg 1$, where $g_X(\tau)$
exhibits an exponential decay $\sim e^{-\pi^2\tau/2}$.
For $\tau = 0$ the series in Eq.~(\ref{COMswarm_result}) can be summed up leading to the variance $\average{X^2} = g_X(0) = 1/(3N)$, which is smaller than the corresponding result \eqref{corr_nocorr} for model A by a factor of 6. In Fig. \ref{fig:fig3}(a) we compare Eq.~(\ref{COMswarm_result}) with simulations and observe a very good agreement.

We now turn to the fluctuations in the system's radius. Averaging Eq.~\eqref{ell_interacting} over the noise, one can see that, within the linear theory, $\average{\delta\ell} = 0$. The two-time autocorrelation of the system's radius is thus given by
\begin{equation} \label{cov_deff_ell}
    g_\ell(\tau) = \average{\ell(0)\ell(\tau)} - \average{\ell}^2 = \average{\delta\ell(0) \delta\ell (\tau)}.
\end{equation}
Plugging Eq.~\eqref{ell_interacting} into Eq.~\eqref{cov_deff_ell} and averaging over the noise we find after some algebra
\begin{eqnarray} \label{CovEll}
    &&g_\ell(\tau) = \frac{2}{\pi N}  \sum_{n = 1}^\infty A_n  e^{-\frac{1}{2} \pi ^2 n^2 \tau }, \\
    &&A_n = \frac{1}{n}\times\begin{cases}
    2 \tanh \left(\frac{\pi  n}{2}\right)+\coth \left(\frac{\pi  n}{2}\right) & n \;\text{odd}, \\
    \tanh \left(\frac{\pi  n}{2}\right) & n \; \text{even}.
    \end{cases}\nonumber
\end{eqnarray}
For $\tau \neq 0$ the sum in Eq.~\eqref{CovEll} converges and agrees well with MC simulations, see Fig. \ref{fig:fig3}(b). However, for $\tau = 0$, the infinite series in Eq.~\eqref{CovEll} diverges logarithmically, because the large-$n$ asymptotic of $A_n$ scales as $1/n$:
\begin{equation} \label{A_n_asymp}
    A_n \simeq \begin{cases}
    3/n , & n \;\text{odd}, \\
    1/n ,& n \; \text{even},
    \end{cases}
\end{equation}
implying an infinite variance of $\ell(t)$. Needless to say, the original microscopic model exhibits a finite variance of the system's radius, as MC simulations  show (see below).
To resolve this contradiction we follow the line of argument of Ref. \cite{siboni2021fluctuations}, where a similar apparent divergence was observed.
We return to Eq.~\eqref{CovEll} and recall that the coarse-grained Langevin description is only valid at macroscopic time lags, $\tau \gg 1/N$. Therefore we can introduce a cutoff at $\tau \sim 1/N$ which yields $g_\ell(0)$ with logarithmic (in $N$) accuracy.

\begin{figure}[t!]
\centering
\includegraphics[width=1.03\linewidth]{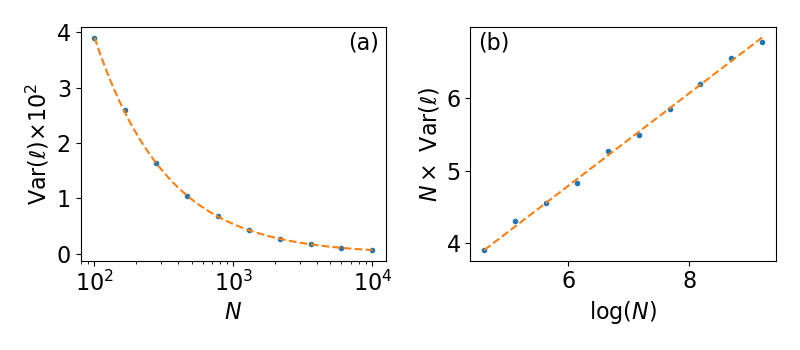}
 \vspace{-7mm}\caption{The variance of the system's radius [model B, Eq.~\eqref{varell_cutoff}] compared with MC simulation (points). In (a) the variance is plotted as a function of $N$. In (b) the variance times $N$ is plotted as a function of $\ln N$. The dashed lines represent the function $N \text{var}(\ell) = (2/\pi)\ln N  + 0.97$, where the factor $0.97$ is fitted and corresponds to a numerical factor under the logarithm. This factor is beyond the logarithmic accuracy of Eq. \eqref{varell_cutoff}.  }
 \label{fig:fig4}
\end{figure}

The calculations proceed in the following way. Using the large $n$ asymptotic of $A_n$ given by~\eqref{A_n_asymp}, we replace the summation in Eq.~\eqref{CovEll} by integration, and approximate the results for $|\tau| \ll 1$, ultimately leading to
\begin{equation} \label{g_ell_approx}
    g_\ell(\tau) = \frac{2}{\pi N}\ln\left(\frac{1}{|\tau|}\right)
\end{equation}
for $1/N \ll |\tau| \ll 1$.  To evaluate the variance, we introduce a logarithmic cutoff in Eq.~\eqref{g_ell_approx} at $\tau = 1/N$, and arrive at the variance with logarithmic accuracy:
\begin{equation} \label{varell_cutoff}
    \text{var} \; \ell(t) \simeq \frac{2}{\pi} \frac{\ln N}{N}.
\end{equation}
The presence of the large logarithmic factor $\ln N$ is noteworthy. Strikingly, Eqs. (\ref{g_ell_approx}) and (\ref{varell_cutoff}) are identical, including the coefficient $2/\pi$, to those obtained for the Brownian bees \cite{siboni2021fluctuations}. In Fig. \ref{fig:fig4} we compare Eq.~(\ref{varell_cutoff}) with MC simulations and observe a good agreement.

\begin{figure}[t!]
\centering
\includegraphics[width=0.75\linewidth]{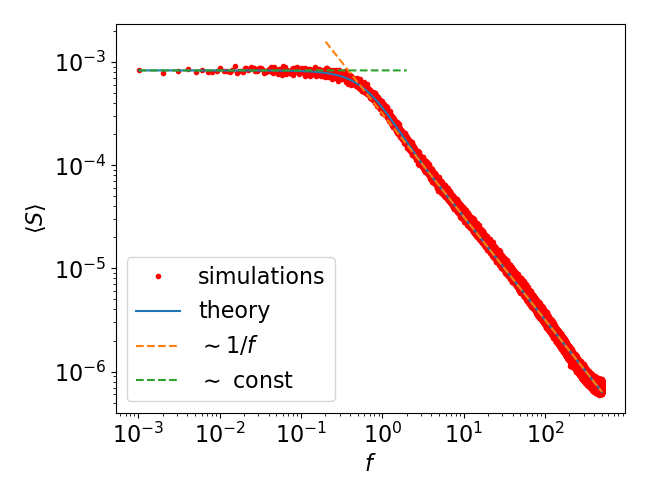}
 \vspace{-3mm}\caption{Theoretical prediction \eqref{PSDfull} for  the power spectral density of the system's radius (solid line) in model B and its asymptotics  \eqref{PSDapprox} (dashed lines) are compared with MC simulation results (red points).  The simulation parameters are $N = 10^3$, the total simulation time is $t = 10^3$, the number of simulations is 500, and the sample rate is $N$.}
 \label{fig:fig4b}
\end{figure}

Notably, in the frequency domain, the logarithmic scaling with $\tau$, described by Eq.~\eqref{g_ell_approx},  corresponds to a $1/f$ noise, as was already observed for the Brownian bees \cite{siboni2021fluctuations}. The $1/f$ noise has been observed in the power spectral density (PSD) of a multitude of stochastic processes \cite{mandelbrot1967some, mandelbrot2002gaussian, handel}. We now briefly describe this connection. For stationary processes the PSD is related to the autocorrelation function, Eq.~\eqref{cov_deff_ell}, by the Wiener-Khinchin theorem:
\begin{equation}
    \average{S(f)} = 2\int_{0}^{\infty} g_\ell(\tau) \cos(2\pi f \tau)  d\tau.
\end{equation}
We thus Fourier transform Eq.~\eqref{CovEll}, resulting in
\begin{equation} \label{PSDfull}
    \average{S(f)} =\frac{8 }{\pi N}  \sum_{n = 1}^\infty  \frac{A_n n^2}{16 f^2 + n^4\pi^2}     ,
\end{equation}
with $A_n$ given by Eq.~\eqref{CovEll}. Two limits are of particular interest here: the low-frequency limit $f \ll 1$ which corresponds to long time lags $\tau \gg 1$, and the high-frequency limit $1 \ll f \ll N$ which corresponds to short (but still macroscopic) times lags $1/N \ll \tau \ll 1$. In each of these cases we Taylor expand Eq.~\eqref{PSDfull}, ultimately leading to
\begin{equation} \label{PSDapprox}
    \average{S(f)} \simeq \frac{1}{N}\times \begin{cases}
    \frac{5}{6}+ \mathcal{O}(f^2) ,&  f \ll 1 , \\
 \frac{1}{\pi f}  + \mathcal{O}(f^{-2}) , & f \gg 1 .
     \end{cases}
\end{equation}
At long times (small $f$) the PSD approaches a constant, consistent with uncorrelated (white) noise, while at short times (large $f$) the PSD exhibits a $1/f$ noise.
In Fig. \ref{fig:fig4b} we compare our theoretical predictions  \eqref{PSDfull} and \eqref{PSDapprox} of the power spectrum of the system's radius with the power spectrum as computed from MC simulations.  A good agreement is observed over a broad range of frequencies.

\section{First passage time to a distant target}
\label{MFPT}

Let us add an additional ingredient to our systems of Brownian particles with reset. Suppose that there is a static target at $x=L$. Upon the first passage of a particle to the target, the process is stopped. What is the MFPT? We start with model A, where the calculation employs the known single-particle results \cite{evans2011diffusion}.

\subsection{MFPT for Model A}

When $L$ is sufficiently large (we will obtain the condition shortly), the expected MFPT to the target at $x=L$ is much longer than the characteristic time, $O(1)$, of establishment of the steady state. In this long-time limit the single-particle survival probability of the target $S(t)$ (the probability not to reach the target until time $t$) is~\cite{evans2011diffusion}
\begin{equation}\label{S1}
S_1(t) \simeq \exp\left(-t e^{-L}\right)\,.
\end{equation}
Correspondingly, the single-particle probability density of arriving at the target at time $t$ is
\begin{equation}\label{f1}
f_1(t) = -\frac{dS(t)}{dt} \simeq  e^{-L} \exp\left(-t e^{-L}\right)\,.
\end{equation}
The probability density of one of $N$ particles to first reach the target is, therefore,
\begin{eqnarray}
  f_N(t) &=& N f_1(t)\, S^{N-1}(t) \nonumber\\
 &\simeq& N e^{-L} \exp\left(-t e^{-L}\right) \exp\left[-(N-1)t e^{-L}\right]\nonumber \\
 &=& N e^{-L} \exp\left(-N t e^{-L}\right)\,.
 \label{fN}
\end{eqnarray}
The MFPT, $\langle T \rangle$, is given by the first moment of $f_N(t)$:
\begin{equation}\label{MFPTA}
\langle T \rangle = \int_0^{\infty} t f_N(t) dt  \,\simeq\, \frac{1}{N}\,e^L\,.
\end{equation}
To be consistent with the long-time asymptotic (\ref{S1}), we must demand that $\langle T \rangle \gg 1$, which leads to the condition  $L\gg \ln N$, that is $L$ must be much larger than the system's radius, see Eq. \eqref{av1}.

\begin{figure}[t!]
\centering
	\includegraphics[width=1.0\linewidth]{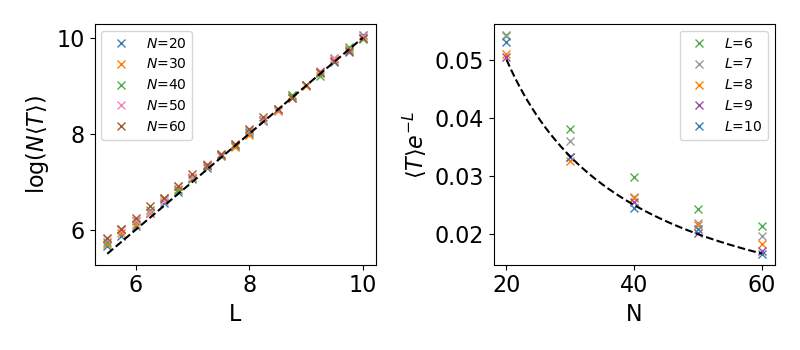}
	\vspace{-5mm}\caption{Rescaled MFPT to the static target at $x=L$ as a function of $L$ (a) and of $N$ (b) for model A. x marks: simulations, dashed line: theoretical prediction (\ref{MFPTA}).}
	\label{fig:mfpta}
\end{figure}
Figure \ref{fig:mfpta} compares the theoretical prediction (\ref{MFPTA}) with simulation results. A very good agreement is observed for sufficiently large $L$.

\subsection{MFPT for Model B and for Brownian bees model}

Due to the inter-particle correlations in these two models, the determination of the MFPT is a hard problem.   Here we  obtain an upper bound for the MFPT for these two models and compare this bound with simulation results for model B. Let us temporarily reintroduce the original variables, where the reset rate is $r$ and the diffusion constant is $D$. For a \emph{single} Brownian particle, resetting to $x = 0$ at rate $r$, the MFPT  to the target at $x=L$ is exactly given by Ref.~\cite{evans2011diffusion}:
\begin{equation} \label{FPT_single}
    \average{T_1} =  \frac{1}{r}\left[\exp\left(\sqrt{\frac{r}{D}}L\right) - 1\right].
\end{equation}
If $L$ is smaller than or comparable with $\ell_0$ (the system's radius of compact support in the hydrodynamic limit), the MFPT of our particle systems will strongly depend on the initial particle positions. Of most interest, therefore, is the limit of sufficiently large $L-\ell_0$, when  $\average{T}$ is expected to be much longer than the characteristic relaxation time $\sim 1/r$ of the system to its NESS.

Our upper bound on $\average{T}$ assumes, both for model B and for the Brownian bees model, a breakaway, or evaporation, scenario. In this scenario, a single particle breaks away from the rest of particles to make an unusually large excursion and reach the target before being reset (to the origin in model B, or to the location of one of the other particles in the Brownian bees model). Typically, particles which start close to $x=\ell_0$, have the highest chance of reaching the target before being reset.  Still most of their attempts to reach the target  fail because of the reset.  After a resetting event, a different particle, also from a close vicinity of $x=\ell_0$, breaks away and attempts to reach the target, etc. The upper bound is obtained when we restrict the ensemble to such particles.    Thus we arrive at an effective \emph{single-particle} process where the effective particle is reset with rate $N r$ to the point $x=\ell_0$. The bound is then given by Eq.~\eqref{FPT_single} with $r$ replaced by $N r$ and $L$ replaced by $L - \ell_0$, resulting in
\begin{equation} \label{FPT_swarm}
    \average{T} \simeq  \frac{1}{N r}\exp\left(\sqrt{\frac{r N}{D}}(L-\ell_0)\right) ,
\end{equation}
where we dropped the term $-1$ inside the square brackets of Eq.~\eqref{FPT_single} to avoid excess of accuracy. Crucially, $\average{T}$ as described by Eq.~(\ref{FPT_swarm}) scales exponentially with $\sqrt{N}$, rather than with $N$. Therefore, the single-particle evaporation scenario is exponentially more efficient than any macroscopic scenario which involves ${\cal O}(N)$ effective number of particles.

We must demand for self-consistency that the MFPT in Eq.~(\ref{FPT_swarm}) be much longer than the relaxation time to the NESS, that is  $\average{T} \gg 1/r$.  Going back to the units where $r=D=1$,
this strong inequality yields
$L-\ell_0 \gg \ln N/\sqrt{N}$.
For both model B and the Brownian bees, this condition coincides, up to a power of $\ln N$, with the condition that $L-\ell_0$ is much larger than the standard deviation of the system radius $\ell(t)$, see Eq.~(\ref{varell_cutoff}).

The evaporation scenario that we adopted here is similar in spirit to the ``eigenvalue evaporation" scenario in  random matrix theory \cite{kosterlitz1,kosterlitz2,parisi1,majumdar,parisi2}. The latter is known to provide \emph{exact} asymptotic results for the statistics of the largest eigenvalues.

How close is the upper bound \eqref{FPT_swarm} to the actual MFPT? To answer this question, we ran stochastic simulations of the first passage in model B. For large $N$ and $L-\ell_0$, direct MC simulations are very costly in terms of simulation time. For this reason we employed  highly efficient \emph{weighted ensemble} (WE) simulations \cite{huber1996weighted, vilk2020extinction},  see the Appendix. As evidenced by Fig. \ref{fig:fig6}, direct and WE simulations give similar results in the parameter regimes which both methods can cover.

\begin{figure}[t!]
\centering
	\includegraphics[width=0.68\linewidth]{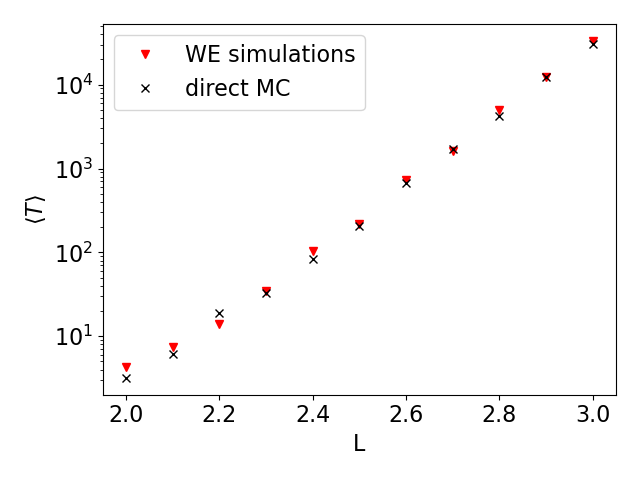}
	\vspace{-5mm}\caption{The MFPT to the static target at $x=L$ as a function of $L$ for model B with $N = 100$, obtained by direct MC simulation (x marks) and WE simulations  (points).}
	\label{fig:fig6}
\end{figure}

\begin{figure}[t!]
\centering
\includegraphics[width=1.03\linewidth]{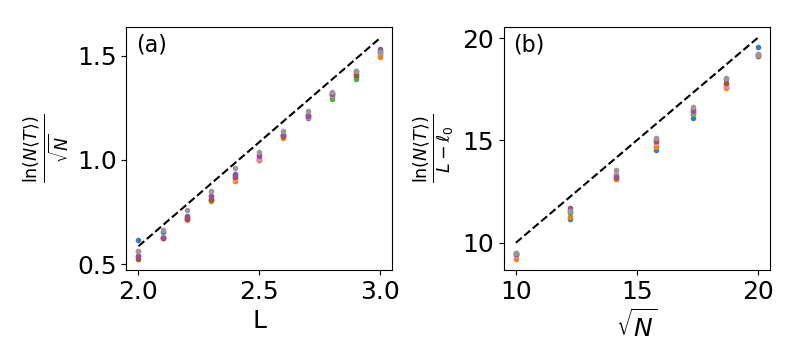}
 \vspace{-5mm}\caption{The MFPT of a particle to $x = L$ for model B as obtained by WE simulations. (a): $\ln (N\average{T})/\sqrt{N}$ as a function of $L$ (dashed line) is compared with the simulation results for different values of $N$ ($100<N<400$) (points). (b):  $\ln (N\average{T})/(L-\ell_0)$ as a function of $\sqrt{N}$ (dashed line) is compared with simulations for different values of $L$ (points).}
 \label{fig:fig5}
\end{figure}

The bound \eqref{FPT_swarm} to the MFPT can be rewritten (in the units of $r=D=1$) as
\begin{equation}\label{logbound}
\ln (N\average{T})\simeq \frac{L-\ell_0}{\sqrt{N}} .
\end{equation}
In Fig. \ref{fig:fig5} we compare this prediction with the WE simulations separately as a function of $L$ and $N$. As to be expected from an upper bound, Eq.~\eqref{FPT_swarm} slightly overestimates the MFPT for all values of $N$. Nonetheless, the functional dependence of $\average{T}$  on $L$ and on $N$ appears to be captured correctly.

A salient feature of Fig.~\ref{fig:fig5} is that the relative accuracy of Eq.~\eqref{FPT_swarm} in its description of the  simulation results does not visibly improve with the increase of $N$. To explain this feature, we notice that the bound \eqref{FPT_swarm} can be improved if one exploits typical steady-state fluctuations of $\ell(t)$ and replaces the effective single-particle reset point $x=\ell_0$ by $x=\ell_0 + a N^{-1/2}$ with some positive numerical factor $a=O(1)$. (Here we ignore the $\ln N$ factor in Eq.~\eqref{varell_cutoff}.) This gives rise to an additive term in the exponent which is independent of $N$.

\section{Discussion}
\label{discussion}

In this work we studied stationary fluctuations in models A and B which involve $N$ Brownian particles subject to stochastic reset in one dimension with different reset rules. We combined exact probabilistic methods  with
a coarse-grained, Langevin-type approach, previously derived for systems of reacting and diffusing particles for $N\gg 1$.

Employing linearization of the Langevin equation around the steady state solutions, obtained in the hydrodynamic limit of $N\to \infty$, we calculated, for both models, the two-time autocorrelation function, and in particular the variance, of the center of mass. Similarly to the previously studied Brownian bees model \cite{siboni2021fluctuations}, the variance of the center of mass for both models scales as $1/N$ as could have been expected from the law of large numbers. The independent character of particles in model A enabled us to verify our macroscopic results in exact microscopic calculations, and to extend the results for the center of mass to its large deviations. In particular, we calculated  the exact distribution of the center of mass for \emph{any} number of particles.

We also studied, in both models,  fluctuations of the system's radius. In model A we determined the exact distribution of this quantity for any $N$. At large $N$, the average system's radius behaves as $\ln N$, whereas the variance of the system's radius is $N$-independent.  In model B the average system's radius is independent of $N$, whereas the variance  scales  as $(1/N) \ln N$.  The unusual latter scaling results from a logarithmic behavior of the two-time autocorrelation function at steady state which, in the frequency domain, corresponds to  $1/f$ noise. These behaviors of model B are shared by the Brownian bees model \cite{siboni2021fluctuations}. This fact hints at universality of this scenario for a whole class
of reset models where only the farthest particles are subject to reset, while the exact destination of the reset particle is irrelevant as long as it is in the system's bulk. This issue is presently under a detailed study \cite{VSS}.
%

Finally, we studied the MFPT to a distant target in models A and B and in the Brownian bees model.
For model A this quantity directly follows from the known single-particle results.
For model B and the Brownian bees model we proposed a sharp upper bound for the MFPT, based on the evaporation scenario similar to one that appears in the random matrix theory. The bound is determined
by a single particle which evaporates from the bulk of the particles to reach the distant target.
Our weighted-ensemble simulations of model B showed that this bound also gives a good approximation to the first-passage time, both as a function of the distance to the target $L$ and total number of particles $N$.

Comparing Eqs.~\eqref{MFPTA} and~\eqref{FPT_swarm} (with $r=D=1$), one can see that in both cases the MFPT to a distant target is exponentially long with respect to $L$ or $L-\ell_0$. In
model A, however,
it is exponentially (in $\sqrt{N}$) shorter than in the two other models. This is not surprising, as the reset rule of model B and of the Brownian bees model discourages the most efficient explorers of distant regions.


\section*{Acknowledgements}
We are very grateful to Satya N. Majumdar for advice and to Pavel Sasorov and Naftali R. Smith for useful discussions. We acknowledge support from the Israel Science Foundation (ISF) through Grants No. 531/20 (OV and MA) and 1499/20 (BM).

\appendix

\section{Weighted Ensemble Simulations}
\label{WEsimulations}

In Sec. \ref{sec:independent} and \ref{sec:InteractingParticles} we compared our analytical results with direct continuous-time Monte Carlo simulations of the microscopic model \cite{gillespie1976general,bhat2016stochastic}. However, direct simulations become prohibitively long for the purpose of determining, in Sec. \ref{MFPT}, the MFPT  to the stationary target. For the latter  we used more efficient weighted ensemble (WE) simulations. The idea of the WE method is to run significantly more simulations in regions of interest, while redistributing the statistical weights of the trajectories accordingly. To this end, space is divided into bins, which can be predefined or interactively chosen (on the fly), to ensure sampling in specific regions of interest. We start the simulation with $m$ ensembles, each with $N$ particles that are located in the origin. Each of the $m$ ensembles are given initial equal weights of $1/m$. The simulation consists of two general steps: (a) ensembles are advanced in time for time $\tau_{WE}$, where the time-propagation method follows the Gillespie algorithm \cite{gillespie1976general, gillespie1977exact}; (b) ensembles are re-sampled as to maintain $m$ trajectories in each occupied bin, while bins that are unoccupied remain such. The process of re-sampling itself can be done in various ways, as long as the distribution is maintained. In our simulation we used the original re-sampling method suggested by Huber and Kim \cite{huber1996weighted, vilk2020extinction}.
Note that $\tau_{WE} \ll 1$ is much shorter than the system's relaxation time, but much longer than the typical time between elemental processes so as to increase efficiency. We also emphasize that bins need to be chosen wisely: if  too far apart, trajectories will not reach remote regions, while if chosen too close together, the computational cost will be very high. Generally, there is a tradeoff between the number of bins and the trajectories per bin, assuming some memory limit. In our simulations, to achieve high efficiency we interactively changed the binning.

We checked that the WE simulations results coincide with ``brute force" MC simulations in parameter regimes where the latter are applicable, see \textit{e.g.}, Fig.~\ref{fig:fig6}. Notably, WE simulations were much more efficient than brute-force MC simulations: while the latter ran for more than two weeks to produce Fig.~\ref{fig:fig6}, the former ran under an hour. We performed error evaluation numerically by running the simulations for different $\tau_{WE}$. The maximum error that we encountered was $15\%$.

\bibliography{references}

\end{document}